# Rapid Mobile App Development for Generative AI Agents on MIT App Inventor

Jaida Gao[1], Calab Su[2], Etai Miller[3], Kevin Lu[4], Yu Meng[5]

*Abstract*—The evolution of Artificial Intelligence (AI) stands as a pivotal force shaping our society, finding applications across diverse domains such as education, sustainability, and safety. Leveraging AI within mobile applications makes it easily accessible to the public, catalyzing its transformative potential. In this paper, we present a methodology for the rapid development of AI agent applications using the development platform provided by MIT App Inventor. To demonstrate its efficacy, we share the development journey of three distinct mobile applications: *SynchroNet* for fostering sustainable communities; *ProductiviTeams* for addressing procrastination; and *iHELP* for enhancing community safety. All three applications seamlessly integrate a spectrum of generative AI features, leveraging OpenAI APIs. Furthermore, we offer insights gleaned from overcoming challenges in integrating diverse tools and AI functionalities, aiming to inspire young developers to join our efforts in building practical AI agent applications.

## I. INTRODUCTION

In the current advancement of the world, artificial intelligence (AI) is the focal point of every aspect of life as every company dives deeper and deeper into this revolutionary technology. AI is founded on the premise that human thought processes can be replicated in physical form. Its origins trace back to the early 1900s, notably with Alan Turing's groundbreaking work [1]. Turing proposed that human cognition emerges from a state of directionless thought, which evolves into coherence through training, ultimately leading to intelligent behavior. He initiated research in Machine Intelligence and introduced the concept of the "Turing Test", a benchmark where a machine's ability to engage in conversation indistinguishable from a human via teleprinter signifies intelligence. The formal introduction of AI occurred at the Dartmouth Workshop in 1956, led by John McCarthy, with the ambitious goal of emulating human behavior.

Despite numerous challenges and setbacks, AI has progressively integrated into everyday life for the past 10 years, from Siri to ChatGPT, from Roombas to self-driving cars [9]. The earlier someone is exposed to AI, the more advantage they will have over their peers in the future as now all attention is on AI. In addition, the one thing most people interact with the most every day is their phone; nowadays globally people average 6 hours and 58 minutes a day of screen time, 3 hours and 43 minutes of that is on their phones; the numbers have been steadily increasing in the past 20 years; however, some studies show a 30% increase of screen time on phones. With the consumer market being so large and the potential for AI to increase, direct, and monitor internet traffic, there is a new subcategory of AI that can be explored. Generative AI is getting incredibly big as we can see already from its implementation on social media by providing users with directed content and by analyzing data much better and faster than traditional algorithms. For application developers, it is imperative to address the critical need for expeditious development of mobile applications for generative AI agents that quickly translate conceptualization into tangible software solutions.

After exploring various methodologies and tools, we have found that MIT App Inventor offers a feasible solution that streamlines the development process and enables efficient transformation of ideas into functional AI agent applications [2, 3, 5, 13]. Over the past two years, our team has developed a diverse array of award-winning AI agent applications, with a focus on sustainability, education, and safety. MIT App Inventor has demonstrated its immense value as an educational platform, providing students at secondary schools with a streamlined introduction to the expansive landscape of computer technology and a frictionless coding experience. Furthermore, we have found that the transition from utilizing MIT App Inventor to professional development environments is seamless and efficient, particularly for proficient students. As we reflect on our journey, we have accumulated numerous valuable insights. Through this paper, our goal is to disseminate these learnings and motivate others, especially young developers, to transform their innovative ideas into tangible AI agent applications with significant societal impact.

## II. RELATED WORK

MIT App Inventor is an online coding tool that allows anyone to build endless apps according to their liking due to its simplicity [4, 6, 8]. As shown in Fig 1, MIT App Inventor operates in the form of block code, this is a form of simple coding that has potential for less errors and less complexity. The user can use pre-programmed blocks that can be dragged together to create a functional app. The visual aspect of coding offered by platforms like MIT App Inventor is particularly advantageous for young learners who are just beginning to explore coding and brim with creative ideas [4, 7, 10]. These visual representations facilitate their understanding of coding theory and the sequence of code actions.

Jaida Gao is the corresponding author (e-mail: jaidagao@gmail.com). [1] is affiliated with San Joaquin Delta College, [2] is affiliated with Trinity Christian Academy, [3] is associated with Venice High School, [4] is a student at St. Mark's School of Texas, [5] is the Co-Founder and CEO of ArcGen Technologies, LLC.

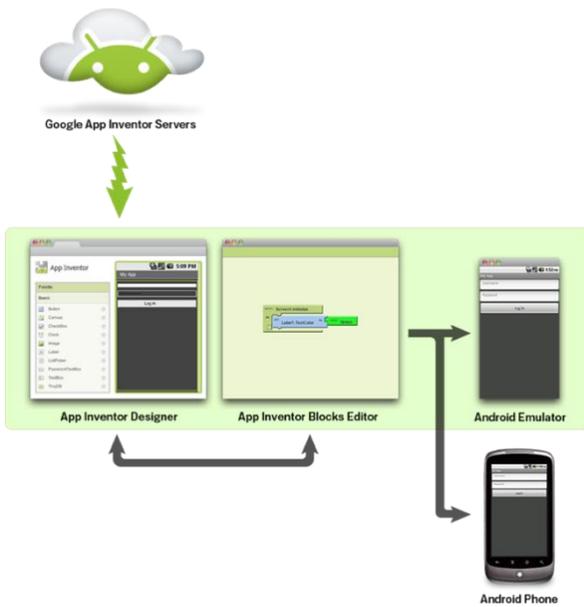

Fig. 1. Architecture of MIT App Inventor [2]

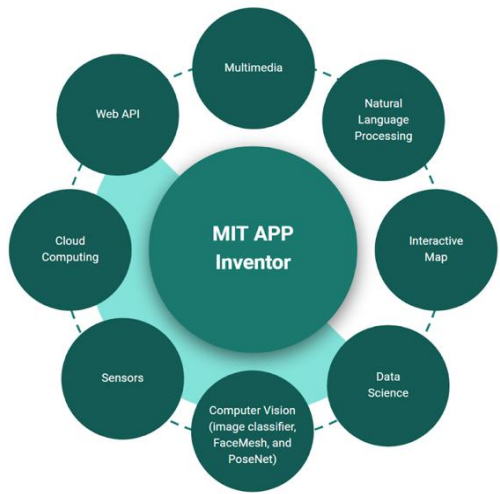

Fig. 2. Potential Functionalities of an MIT App Inventor App

As shown in Fig 2, MIT App Inventor stands as a robust and flexible development platform, uniquely adaptable across a broad spectrum of applications [14, 15]. Its seamless integration with cloud-based services via Web APIs and HTTP requests opens the door to leveraging the capabilities of advanced tools, including OpenAI's suite of AI technologies. Utilizing the tools, developers can easily create conversational AI agents like Alexa and Siri; integrate voice commands and image recognition capabilities; or even build personalized chatbots [11, 12]. This feature-rich platform further extends its connectivity by interfacing with a wide array of microprocessors through Bluetooth, Wi-Fi, or Node-Red protocols, showcasing unparalleled versatility.

The true strength of MIT App Inventor lies in its capacity to morph according to the project's needs. Whether functioning as a sophisticated standalone application, an interactive data dashboard, a comprehensive control panel, or a cutting-edge edge device within the burgeoning Internet of Things (IoT) ecosystem, it provides a foundational toolset that encourages creativity and innovation. Its inherent flexibility and ease of use enable developers to craft compelling, multifaceted solutions that span various domains, from education and health to environmental sustainability and beyond. In doing so, MIT App Inventor empowers creators to bridge the gap between complex technological concepts and practical, real-world applications, driving forward the next wave of digital innovation.

Furthermore, mobile devices have become ubiquitous, captivating an ever-expanding user base. These handheld marvels serve as an excellent playground for students to refine their implementation skills. Within the domain of MIT App Inventor, a powerful development environment, lies a wealth of sensors waiting to be explored. Mobile devices come equipped with a variety of sensors that enhance their functionality and enable a wide range of applications. Mobile devices, coupled with MIT App Inventor's sensor-rich environment, provide an exciting canvas for students to unleash their creativity and build practical solutions [16].

## III. METHODOLOGY

In this section, we delve into the systematic methodology utilized for the rapid development of AI agent applications with MIT App Inventor. As shown in Fig 3, the whole process comprises six consecutive stages, each playing a pivotal role in ensuring the overall success of the development process. It starts with ideation and survey to generate ideas and devise an action plan for addressing the problem. Subsequently, the design phase ensues, involving research and further refinement of the plan. The development phase follows, wherein the actual app is constructed. Testing then takes place to gather feedback and identify areas for improvement. These four phases form a continuous cycle; if issues arise in one phase, there's flexibility to revisit previous stages. Upon completion of testing, the evaluation phase begins, drawing conclusions and reviewing achievements. The app's launch marks the deployment of the solution to the public, extending its availability beyond creators and testers. Continuous feedback and data collection during this phase facilitate ongoing development in the future. The cycle repeats, whether it involves refining the existing project or initiating a new one.

Next, we will provide a detailed explanation of each phase.

1) The ideation and survey phase serves as a foundational step in shaping the structure, features, and design concepts of the mobile application. Through iterative brainstorming sessions, factors such as user feedback, market dynamics, and technological advancements are carefully considered. This stage recognizes the possibility of revisiting the research phase for additional refinement as needed.

2) The design phase involves translating planned concepts into a visual and functional blueprint, with a focus on user-centric design principles. MIT App Inventor's visual programming environment is leveraged to develop wireframes, prototypes, and user interfaces. The aim is to ensure coherence with the design specifications outlined during the planning phase.

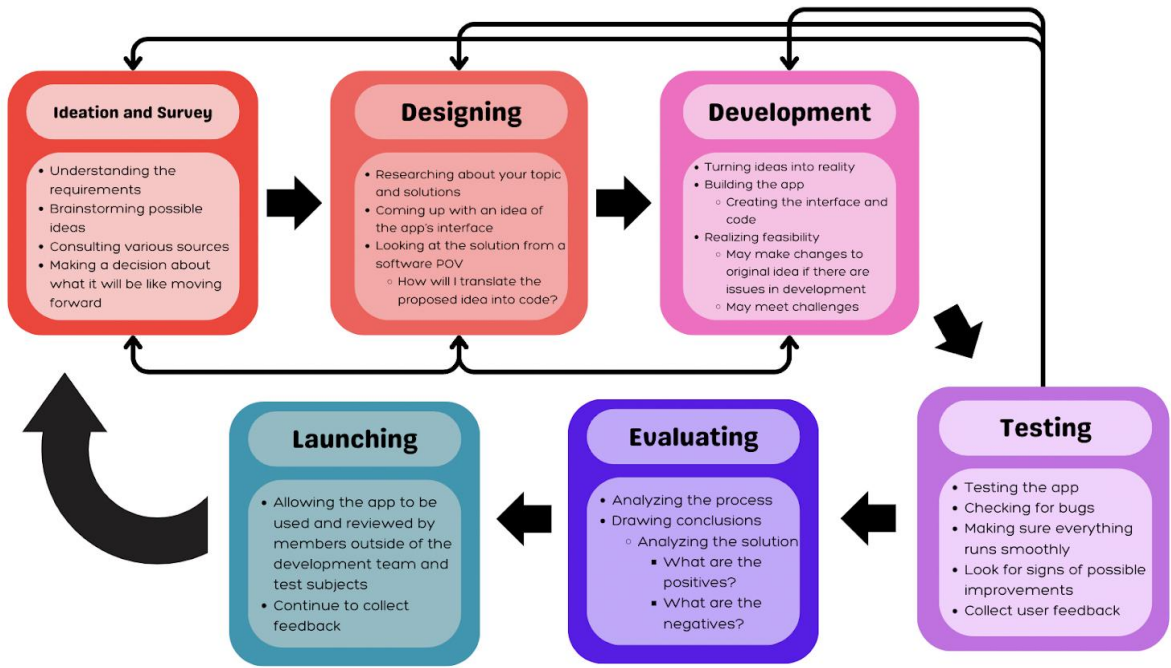

Fig. 3. Iterative Process of AI Agent Development

3) In the development phase, planned features are implemented, and AI functionalities are integrated, utilizing MIT App Inventor's modular capabilities. This modular approach accelerates development and improves maintainability, while the incorporation of Web APIs facilitates seamless integration of generative AI features. Maintaining alignment with design specifications remains a primary focus throughout this phase.

4) Thorough testing is carried out to verify the stability and functionality of the developed application. This includes unit testing, integration testing, and user acceptance testing. In case any issues are identified, an iterative loop is initiated, enabling a return to the development phase to address challenges and ensure alignment with design specifications.

5) After testing, the application's performance is reviewed, incorporating testing feedback for ongoing improvement.

6) The deployment/launch phase entails releasing the developed application to the intended platform, emphasizing a seamless deployment process. Standard deployment procedures are followed, and the application's performance post-launch is closely monitored to promptly address any potential issues. An iterative approach is employed afterwards, enabling revisits to the design, development, or testing phases as needed. This iterative process enables adaptation to evolving requirements and the seamless integration of user-driven enhancements.

This comprehensive methodology offers a structured framework for rapidly developing AI agent applications, ensuring technological robustness while staying responsive to user needs and market dynamics. The iterative nature of the methodology enables continuous refinement, thereby enhancing the overall success of the applications. Finally, meticulous documentation is diligently upheld throughout the entire development

```python
# Python Code: Calculate the Fibonacci sequence using a while loop
n = 10
fibonacci_sequence = []
a, b = 0, 1
count = 0
while count < n:
    fibonacci_sequence.append(a)
    a, b = b, a + b
    count += 1
print("Fibonacci Sequence:", fibonacci_sequence)
```

Fig. 4. A Python Code Snippet and its Corresponding MIT App Inventor Block Code

process, documenting design choices, implementation specifics, AI model integrations, testing outcomes, and other pertinent details. This documentation acts as an asset for future development endeavors, fostering transparency and guaranteeing the reproducibility of the research.

TABLE I.  Block Coding vs. Line Coding

|  | Block Coding (MIT App Inventor) | Line Coding e.g. Python, C/C++ |
|---|---|---|
| **Coding Environment** | Simple UI; tools are conveniently located in a menu. | Detailed workspace that can cause stress to a new coder. Tools are hidden in many different places. |
| **Learning Complexity** | Easy learning curve, straightforward drag and drop. | More difficult to learn and even harder to master. Very technical and time consuming. |
| **Practicality** | Practical as API, LLM, Computer vision, cloud computing all are easily incorporated through the pre-programed blocks. However, it is more restrictive in the block coding format. | Very practical as it allows more freedom in application. However, it is very complex as the user writes their own functions. |
| **Impact to CS education** | Provides a frictionless learning experience to students at secondary schools | Hard on the technical side, focuses more on implementation. Therefore, it is harder on beginners. |

## IV.  DESIGN, IMPLEMENTATION, AND CASE STUDIES

Over the past two years, our team has developed a range of AI agent applications, with a primary focus on sustainability, education, and safety. We chose to utilize MIT App Inventor due to its user-friendly coding environment and extensive array of features, facilitating diverse interface design and software development. The app development was done primarily in block code, which corresponds to line code, with "if-then" blocks and "dictionary" blocks to name a few. Fig 4 shows a python code snippet and its corresponding MIT App Inventor block code. To illustrate the advantages of MIT App Inventor, Table I provides a comparison between block coding and line coding. MIT App Inventor's rich feature set provides flexibility across diverse functions and niches within app development. While inherently flexible, these features particularly shine in modules tailored for specific domains such as sustainability, education, and safety.

TABLE II.  Expected Features of a Sustainability App

| Feature | Description |
|---|---|
| **Map** | Allows users to see posts near them and view sustainability issues |
| **Your Posts** | Allows the users to view their own posts |
| **Post** | Allows users to create posts for cleanups, alert others of littering, etc. |
| **Main Page** | Users can view the posts near them and reach out to the original posters |

TABLE III.  Expected Features of an Education App

| Feature | Description |
|---|---|
| **Notecards Tool** | A classic tool to consider implementing in Education apps. Tried-and-true method of studying, and convenient to implement using lists and dictionaries. |
| **Diagnostics** | An optimal way to gauge the user's skill level, ideal studying methods, and more using simple tests. While never 100% accurate, it provides an appealing level of personalization. |
| **Calendar** | The classic planning tool, useful for keeping track of deadlines and events. Allows for a versatile Education app, and can be made using CloudDB, lists and dictionaries. |
| **Interpersonal Communicator** | Studies have proven that group work and studies are a highly efficient method to get things done, and by adding social media-esque features to your education app, you can add this appeal to a wider audience. |
| **AI Study Tool** | One of the most versatile features possible in an education app. Easily achievable using the ChatBot feature, allows users to generate many sample problems to assist with their studies. |
| **Timers** | A super easy-to-implement feature that greatly improves the user experience for the app. Using the clock feature, users can now budget their time for studying and other work. |

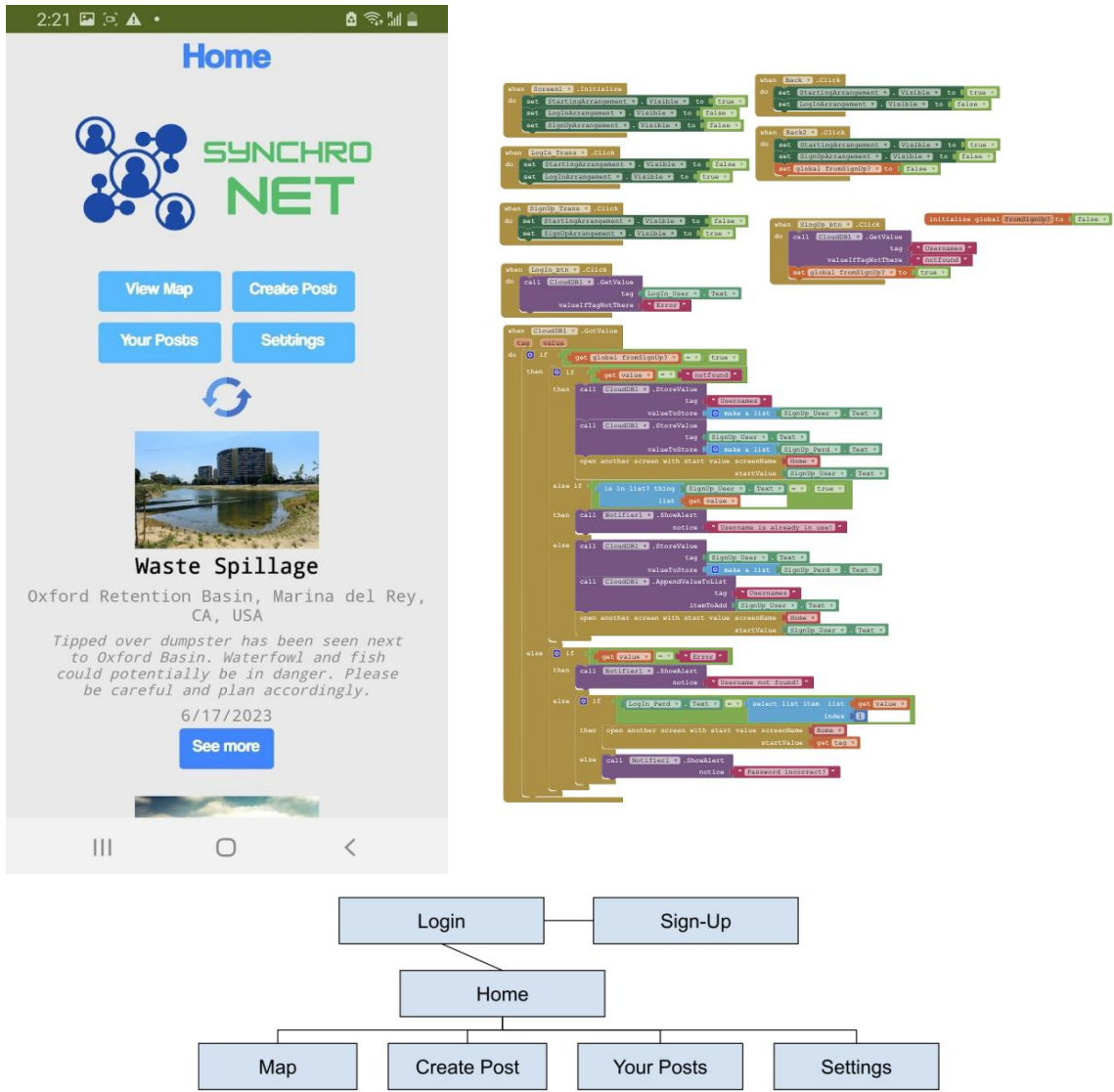

Fig. 5.  Page Design and Block Code for the *SynchroNet* App

### A. App Development for Sustainability

Developing a sustainability-focused app with MIT App Inventor relies on fostering community support and providing incentives for individuals to adopt sustainable practices in their daily lives. Table II lists a set of key features expected by the users of sustainability apps. By harnessing AI and machine learning technologies, these apps should effectively identify environmental hazards and facilitate actions such as recycling, thus empowering users to make tangible contributions to environmental preservation. Machine learning algorithms play a crucial role in tasks such as distinguishing between recyclable materials like plastic, metal, and paper, or evaluating the sustainability of food products based on nutrition labels. Furthermore, Generative AI capabilities enable the identification of environmental threats within a user's community, such as polluted beaches, oil spills, or chemical leaks. These functionalities serve as a catalyst for group collaboration, making it easier for individuals to collectively enhance the sustainability of their communities and contribute to a greener planet. When developing an app focused on sustainability, fostering user engagement is crucial. Sustainability cannot be solely attained through the app; instead, it serves as a tool to empower users with the knowledge of what actions are necessary. Users themselves drive change, and thus the app must offer incentives to encourage their involvement.

MIT App Inventor provides robust support for integrating machine learning models and AI platforms like ChatGPT, as well as connections to messaging and phone applications. This enables app users to adopt sustainable actions more effectively and fosters a sense of achievability in their sustainability goals. As depicted in Fig 5, our team developed the *SynchroNet* app with the objective of fostering sustainable communities. Upon opening the mobile app, users are prompted to either log in or sign up using a username and password. Upon successful authentication, users are directed to the home page, featuring three random posts and navigation buttons to access other app sections. A pivotal feature of the app is an interactive map, accessible via the "*View Map*" button, which displays the location of a random post in the user's vicinity. By clicking the "*Navigate*" button, users can estimate travel time and distance to the selected post location. Additionally, users can create their own posts by clicking the "*Create Post*" button on the home page. After selecting an image and providing necessary information, the post is saved to the *SynchroNet* database, and users are redirected to the "*Your Posts*" page to manage their posts. For assistance in identifying hazards or seeking sustainability tips, users can utilize the AI chatbot feature. Finally, the settings page enables users to log out or change their password as needed.

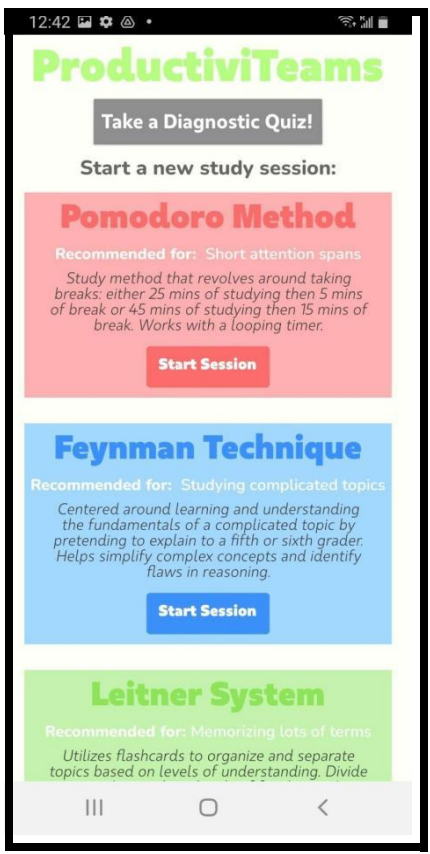

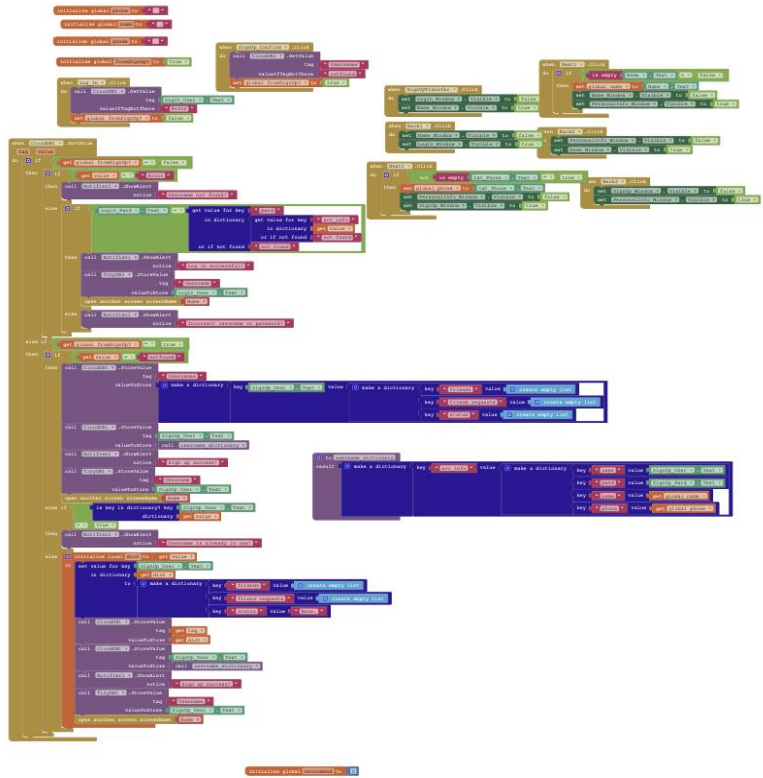

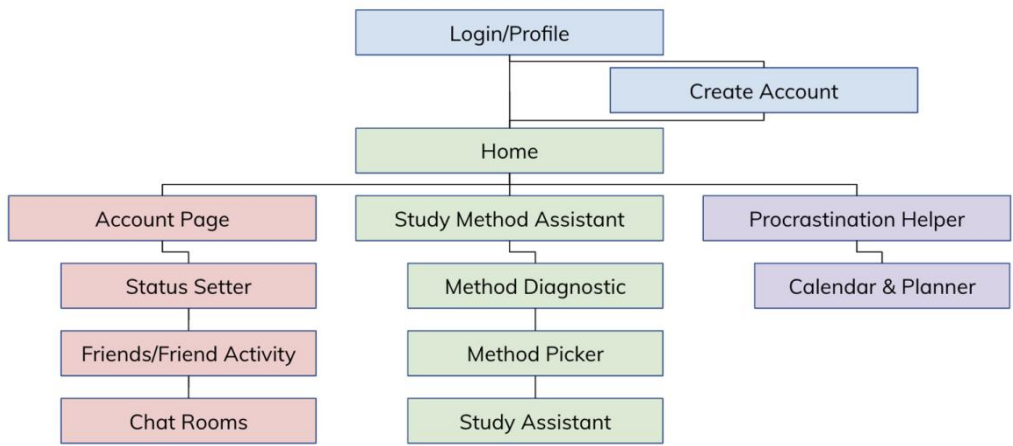

Fig. 6. Design and Implementation for the *ProductivTeams* App

*B. App Development for Education*

When developing an educational app with MIT App Inventor, the focus lies on features that promote good study habits and facilitate information retention, as shown in Table III. AI plays an important role in simplifying many of these tasks, offering flexibility in answering questions. By leveraging ChatGPT and transforming it into a help center through a series of prompts, an effective study tool applicable to any subject can be created. Tools like CloudDB allow the app to store study information for future reference. Additionally, the educational app consists of multiple modules tailored to various studying techniques based on user feedback. Since study techniques vary for each individual, user input becomes invaluable in refining the app's functionality and maximizing its utility. AI's adaptability in iterative design ensures swift improvements to benefit a wider user base when using educational apps.

Fig 6 illustrates the design and implementation of the *ProductivTeams* app, designed to combat procrastination. Like most mobile apps, users begin by creating an account, providing details such as their name, phone number, username, and password. Upon signing up, users are directed to the home page, where they can complete a quick diagnostic quiz to determine their optimal study method. Options include the Pomodoro Method, utilizing a looping timer to maintain focus with intermittent breaks; the Feynman Technique, which involves explaining topics as if teaching a novice; and the Leitner System, a flashcard-based method for spaced repetition. Our app streamlines the process of creating, organizing, and studying index cards, providing a comprehensive study aid. To address procrastination, we've implemented a social-media-like friending system, enabling users to connect and support one another through chat rooms and status updates. The calendar feature assists users in managing deadlines and scheduling study sessions. Additionally, the ChatBot module generates practice questions across various subjects. Finally, users can manage their account settings or log out on the profile page.

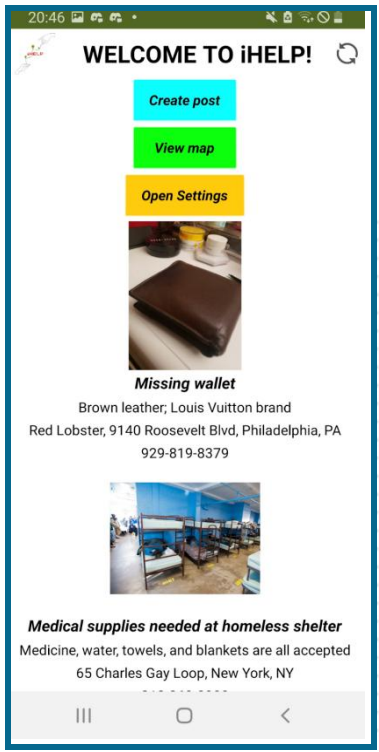

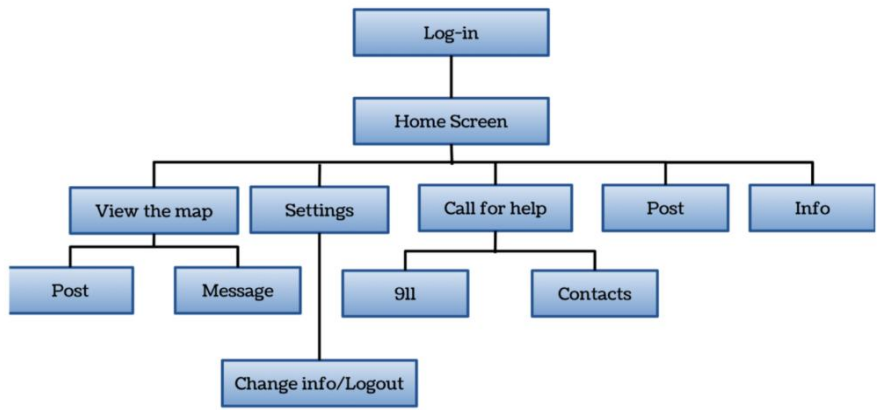

Fig. 7. Design and Implementation for the *iHELP* App

### C. App Development for Safety

Safety stands on its own within app development modules as one where ethical concern takes priority. When using AI integrations to help assess various safety-related subjects, such as building a safe community, it is essential the users know how their data is being handled and who it may be seen by. With features listed in Table IV, safety apps ensure that users are able to get accessible help without being otherwise put in more danger. Using a database like TinyDB to store user data is one of the most private ways to store user data, as it remains on the device the user is on, but with more privacy, functionality is diminished. When using more adaptable and useful functions through Web APIs that utilize AI with increased precision, user data is stored online and is more susceptible to falling into the hands of others. It is of utmost importance to let the user know that their data is being stored and used privately and ethically, in order to have the greatest user satisfaction and trust in a product.

Using MIT App Inventor, we developed the *iHELP* App, an all-in-one app for a healthier and safer community. Fig 7 illustrates its design and implementation. The app was created because we saw many articles on people getting lost and having no way of alerting others. If they did try to contact others, they would have to open multiple apps just to notify others of their situation. Sometimes, there might be no one in your area you can reach out to as well, and finding someone takes a lot of effort. To help those in need, we developed the *iHELP* app with an alert system within a community in one app. Some main features of the app include the navigation map that shows users' requests for assistance as well as police stations and hospitals where they can receive help, a mutual assistance network that allows people give/receive assistance to/from each other, and an info page that shows what to do in case of emergency. Now, people don't have to open up multiple apps or try to find someone who is able to help them. The CloudDB is used to store all users' login info, posts, etc. In addition, the app also has a location sensor and a Google Maps API that can gather and show your location along with where help is needed, as well as a help page where users can easily contact authorities or find someone in their contacts to help them when needed.

TABLE IV.  Expected Features of a Safety App

| Feature | Description |
| --- | --- |
| **Map** | Allows users to see posts near them and communicate with other users in their area |
| **Call for help** | Allows the users to call either the police or a contact for help |
| **Post** | Allows users to create posts for danger, missing items, etc. |

### D. Lessons Learned

Throughout the app development process, our team learned how to work as a team to overcome challenges. There were some differing ideas/opinions at first about the app, but we learned to compromise and create an app that everyone is satisfied with. If someone was having trouble solving an issue with the code, someone else could step in and help, or direct them to other resources. Another thing we learned is how to use MIT App Inventor to create a functional app that can solve a common problem. We learned how to use different databases to store information, as well as create an online community platform where people can easily ask/give assistance.

There were many challenges we faced during the making of this app. The biggest issue we faced was figuring out how to carry out our ideas. There were some functionalities we weren't sure how to implement at first, but we asked our peers, the internet, and teachers. Eventually, we were able to conquer these challenges and it was a very rewarding experience. Additional challenges we faced during the implementation process mainly dealt with integrating the AI features. We found that MIT App Inventor did have some limits when it came to quotas. When we were implementing the randomized questions feature, there was a 10-question quota that couldn't be bypassed. However, this could be solved by reloading the app and going back to the screen, which resets the quota.

## V. CONCLUSION AND FUTURE WORK

In this paper, we illustrate the rapid development of AI agent applications using MIT App Inventor and assorted tools by detailing the creation of three applications aimed at diverse fields. Through this exploration, we underscore the potential of integrating AI into app development to enhance accessibility for the general public.

Moving forward, our efforts will focus on further harnessing the capabilities of MIT App Inventor, incorporating additional AI features, and exploring its potential for controlling virous IoT devices. Additionally, we remain committed to actively participating in local STEAM initiatives and conducting training sessions to inspire and empower young developers to embark on creating impactful mobile applications.